\documentstyle[prb,aps,twocolumn,epsfig]{revtex}
\newcommand{\be}{\begin{eqnarray}}
\newcommand{\ee}{\end{eqnarray}}

\begin{document}
\draft
\twocolumn[\hsize\textwidth\columnwidth\hsize\csname @twocolumnfalse\endcsname

\title{Optical Evidence of Multiphase Coexistence in Single Crystalline
(La,Pr,Ca)MnO$_{\text{3}}$}
\author{H. J. Lee, K. H. Kim, M. W. Kim, and T. W. Noh}
\address{School of Physics and Research Center for Oxide Electronics, Seoul National\\
University, Seoul 151-747, Korea}
\author{B. G. Kim, T. Y. Koo, and S.-W. Cheong}
\address{Department of Physics and Astronomy, Rutgers University, Piscataway, New\\
Jersey 08854}
\author{Y. J. Wang, and X. Wei}
\address{National High Magnetic Field Laboratory, Florida State University,\\
Tallahassee, Florida 32310}
\maketitle

\begin{abstract}
We investigated temperature ({\it T})- and magnetic field-dependent optical
conductivity spectra $\sigma (\omega )$ of a La$_{\text{5/8-}y}$Pr$_{y}$Ca$_{%
\text{3/8}}$MnO$_{\text{3}}$ ({\it y}$\approx $0.35) single crystal, showing
intriguing phase coexistence at low {\it T}. At {\it T}$_{\text{C}}$%
\mbox{$<$}%
{\it T}%
\mbox{$<$}%
{\it T}$_{\text{CO}}$, a dominant charge-ordered phase produces a large
optical gap energy of $\sim $0.4 eV. At {\it T}%
\mbox{$<$}%
{\it T}$_{\text{C}}$, at least two absorption bands newly emerge below 0.4
eV. Analyses of $\sigma (\omega )$ indicate that the new bands should be
attributed to a ferromagnetic metallic and a charge-disordered phase that
coexist with the charge-ordered phase. This optical study clearly shows that
La$_{\text{5/8}{\it -y}}$Pr$_{{\it y}}$Ca$_{\text{3/8}}$MnO$_{\text{3}}$ (%
{\it y}$\approx $0.35) is composed of multiphases that might have different
lattice strains.
\end{abstract}


\vskip1pc]

\newpage

\section{INTRODUCTION}

Recently, electronic phase separation in manganites has been the subject of
various theoretical\cite{Dagotto,Littlewood} and experimental\cite
{Uehara,KHKim,Babushikina,Moritomo,Allodi,Kimura} studies. Computational
studies predicted that a strong tendency toward electronic phase separation
in manganites could lead to an inhomogeneous ground state with hole-rich and
hole-poor regions.\cite{Dagotto} Indeed, many experimental tools probed
evidences of mixed phases with various length and time scales in the wide
phase space of manganites that was obtained by tuning of chemical pressure 
\cite{Uehara,KHKim,Babushikina,Moritomo} and/or carrier (or impurity)
doping. \cite{Moritomo,Allodi,Kimura}

La$_{\text{5/8}-y}$Pr$_{y}$Ca$_{\text{3/8}}$MnO$_{\text{3}}$ revealed a
unique type of phase separation into submicrometer-sized mixtures of
ferromagnetic (FM) metallic and charge-ordered (CO) insulating domains.\cite
{Uehara} With variation of {\it y}, the chemical pressure was varied to tune
the volume fraction and the domain size of each component. It was found that
the metal-insulator transition in the compound occurs through a percolation
process and that magnetoresistance can be dramatically enhanced due to the
two-phase coexistence.

In spite of such interesting phenomena, the physical origin of the phase
separation in La$_{\text{5/8}-y}$Pr$_{y}$Ca$_{\text{3/8}}$MnO$_{\text{3}}$
is not yet fully understood. A hole segregation-type phase separation \cite
{Goodenough} was ruled out because the penalty in electrostatic energy was
too large for the large-scale charge separation of submicrometer-sized
domains.\cite{Littlewood,Uehara} Instead, Littlewood suggested that
structural inhomogeneities should exist due to a large strain mismatch of
the FM metallic and the CO insulating domains.\cite{Littlewood} Related to
this suggestion, a neutron scattering study by Radaelli {\it et al.} showed
that a phase separation with the mesoscopic length scale (500-2000 \AA )
occurring in Pr$_{0.7}$Ca$_{0.3}$MnO$_{3}$ could be driven by intragranular
strain.\cite{Radaelli} On the other hand, through an X-ray scattering study,
Kiryukhin {\it et al.} indicated that an additional insulating phase might
be present in a La$_{\text{5/8}-y}$Pr$_{y}$Ca$_{\text{3/8}}$MnO$_{\text{3}}$
crystal as a function of temperature ({\it T}).\cite{Kiryukhin}

Up to this point, there have been a limited number of optical studies on
phase separation behavior in manganites. Liu, Cooper, and Cheong\cite{Liu}
observed two absorption bands in optical conductivity spectra $\sigma
(\omega )$ of Bi$_{1-x}$Ca$_{x}$MnO$_{3}$ at {\it T}$_{\text{N}}$%
\mbox{$<$}%
{\it T}%
\mbox{$<$}%
{\it T}$_{\text{CO}}$ that could be attributed to $\sim $100 \AA\ scale
mixtures of FM metallic and CO insulating domains. Jung {\it et al}.\cite
{Jung} also observed two infrared (IR) absorption bands in $\sigma (\omega )$
of La$_{\text{1/8}}$Sr$_{\text{7/8}}$MnO$_{\text{3}}$ that suggested a phase
separation between two FM phases which were orbital-ordered insulating and
orbital-disordered metallic phases. These studies addressed the coexistence
of two kinds of phases based on two absorption bands in $\sigma (\omega )$.
However, there has been no report showing an evidence of multiphase
coexistence through optical conductivity studies of manganites.

In this report, we present temperature ({\it T)}- and magnetic field ({\it H)%
}-dependent $\sigma (\omega )$ of a La$_{\text{5/8}-y}$Pr$_{y}$Ca$_{\text{3/8%
}}$MnO$_{\text{3}}$ ($y\approx $0.35) (LPCMO) single crystal. At {\it T}$_{%
\text{C}}$%
\mbox{$<$}%
{\it T}%
\mbox{$<$}%
{\it T}$_{\text{CO}}$, the CO phase shows a large optical gap energy of $%
\sim $0.4 eV, which remains finite above {\it T}$_{\text{CO}}$. At {\it T}%
\mbox{$<$}%
{\it T}$_{\text{C}}$, at least two additional absorption bands newly emerge
below 0.4 eV. The new absorption bands can be attributed to a FM metallic
and a charge-disordered phase, appearing in the backbone of the
charge-ordered phase below {\it T}$_{\text{C}}$. The {\it T}-dependent
changes in the phonon spectra suggest that these multiphases have different
degrees of lattice strain in each

\begin{figure}[tbp]
\epsfig{file=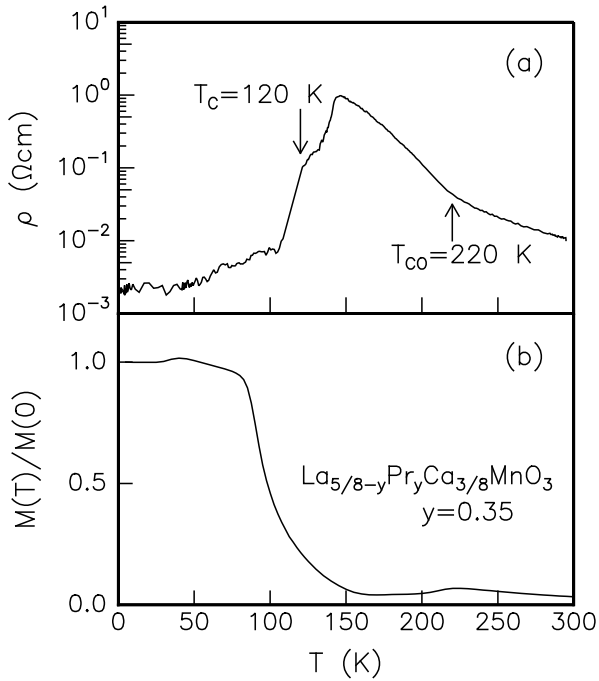, width=3.1in,clip=}
\vspace{2mm}
\caption{(a) $T$-dependent resistivity $\protect\rho $ at zero magnetic
field for a La$_{5/8-y}$Pr$_{y}$Ca$_{3/8}$MnO$_{3}$ ($y$=0.35) single
crystal. (b) $T$-dependent normalized magnetization.}
\label{Fig:1}
\end{figure}

phase. The $H$-dependent $\sigma (\omega )$
indicate that, with increasing {\it H} at 4.2 K, these coexisting
multiphases become unified and eventually turn into a homogeneous FM
metallic phase at 12 T.

\section{EXPERIMENTS AND RESULTS}

\subsection{Transport and magnetic properties}

A single crystal of LPCMO was grown by the floating-zone method using a
mirror furnace. The sample was characterized by resistivity $\rho $ and
magnetization measurements using the four-probe method and a SQUID
magnetometer, respectively. Figure 1(a) shows the {\it T}-dependent $\rho $
data for the LPCMO. With decreasing {\it T}, this sample undergoes a
charge-ordering transition at {\it T}$_{\text{CO}}\approx $220 K and then a
relatively sharp insulator-metal transition around {\it T}$_{\text{C}%
}\approx $120 K. Near {\it T}$_{\text{C}}$, we can observe the steplike
changes of $\rho $ in the metal-insulator transition region. [Even for
samples obtained from the same batch, the temperature where the steplike
structures appear varies.] Similar steplike behaviors were observed not in
LPCMO polycrystals but in single crystals,\cite{podzorov} which was
attributed to the larger size of the FM domains in the single crystals.
Below {\it T}$_{\text{C}}$, the $\rho $ data show some fluctuations, which
seem to be related to the reported random telegraph noise of LPCMO.\cite
{telegraph} Therefore, our $\rho $ response at low {\it T} suggests the
coexistence of FM metallic and CO insulating domains with a temporal
fluctuation. Figure \ 1(b) shows the normalized magnetization

\begin{figure}[tbp]
\epsfig{file=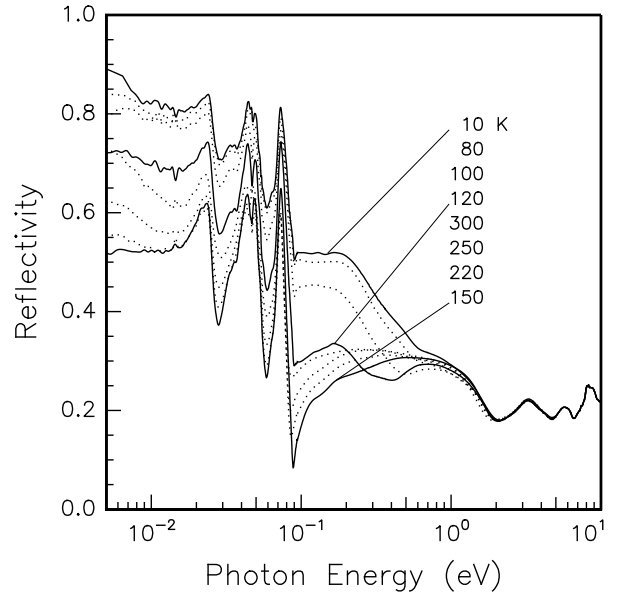, width=3.1in,clip=}
\vspace{2mm}
\caption{(a) $T$-dependent R($\protect\omega $) of a La$_{5/8-y}$Pr$_{y}$Ca$%
_{3/8}$MnO$_{3}$ ($y$=0.35) single crystal. }
\label{Fig:2}
\end{figure}

value, M({\it T%
})/M(0), of the LPCMO crystal. M({\it T})/M(0) increases gradually around 
{\it T}$_{\text{C}}$, which is distinguished with the behavior of{\em \ }the
FM transition in a homogeneous system. And, the value of M({\it T})/M(0)
saturates at low temperature where the FM state becomes dominant.

\subsection{{\it T}-dependence of optical spectra}

We measured reflectivity spectra $R(\omega )$ with various {\it T}. Detailed
techniques for the $R(\omega )$ measurements were described in our previous
report.\cite{HJLee} Figure 2 shows {\it T}-dependent $R(\omega )$ of the
LPCMO single crystal. There are sharp structures due to transverse optic
phonon modes in the far-infrared region. As {\it T} approaches 150 K from
300 K, $R(\omega )$ below 0.5 eV decrease, which is consistent with the dc
resistivity behavior shown in the Fig. 1(a). As {\it T} decreases further
below {\it T}$_{\text{C}}$, $R(\omega )$ show drastic increases, approaching
the metallic response at 10 K.

Using the Kramers-Kronig (KK) relation, we obtained $\sigma (\omega )$ from
the measured $R(\omega )$.\cite{HJLee} Figure 3(a) shows {\it T}-dependent $%
\sigma (\omega )$ above {\it T}$_{\text{C}}$. As {\it T} is lowered from 300
K, $\sigma (\omega )$ below 0.5 eV are systematically suppressed and an
optical gap clearly developed. With decreasing {\it T} and crossing {\it T}$%
_{\text{CO}}$, the peak position of the absorption band near 1.3 eV moves to
higher energy around 1.4 eV. This behavior illustrates that the charge
ordering evolves as a consequence of the freezing out of charge and lattice,
and eventually makes a clean charge gap{\em \ }larger{\em \ }in accord with
the more insulating nature of the CO state. Therefore, it is likely that the
CO phase is dominant at {\it T}$_{\text{C}}$%
\mbox{$<$}%
{\it T}%
\mbox{$<$}%
{\it T}$_{\text{CO}}$ and that the broad band around 1.4 eV can be
attributed to the characteristic optical response of the CO domains.

Figure 3(b) shows that,{\em \ }below {\it T}$_{\text{C}}$, new absorption
bands appear in the low energy region. The features below $\sim $0.5 eV grow
in strength as {\it T} decreases. Note that there is no Drude-like peak even
at the metallic state. As indicated by the asterisks, $\sigma (\omega )$ at
10 K show at least two mid-IR absorption bands centered around 0.22 and 0.49
eV. As {\it T} decreases, the former remains located nearly at the same
frequency, while the latter shifts to a higher frequency from 0.35 eV at 120
K to 0.49 eV at 10 K. This suggests that the origin of the lower frequency
peak might be different from that of the higher frequency one. Even at {\it T%
}%
\mbox{$<$}%
\mbox{$<$}%
{\it T}$_{\text{C}}$, the strength of a broad absorption band around 1.4 eV
does not decrease. This is in contrast with the $\sigma (\omega )$ behaviors
of homogeneous FM metallic samples that show a significant spectral weight
transfer from above 1.0 to below 1.0 eV.\cite{KHKim98} [A similar spectral
weight change was observed by applying $H$, which will be shown in Sec. II.
C.] This suggests that the volume fraction of the CO phase does not change
significantly below {\it T}$_{\text{C}}$.

The evidence of the CO phase below {\it T}$_{\text{C}}$ can also be seen in
the {\it T}-dependent phonon spectra, shown in Fig. 4. Depending on the
types of collective motions in perovskite materials, the phonons around 182,
346, and 572 cm$^{-1}$ are known as external, bending, and stretching modes,
respectively.\cite{KHKim96} Each phonon mode reflects the motion of related
ions: The external mode represents the vibrating motion of the La(Pr or Ca)
ions against the MnO$_{6}$ octahedra. The bending mode is strongly affected
by a change in the Mn-O-Mn bond angle. The stretching

\begin{figure}[tbp]
\epsfig{file=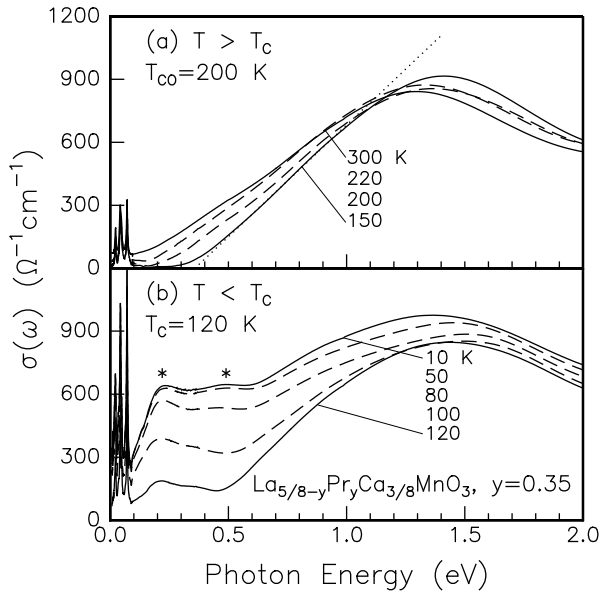, width=3.1in,clip=}
\vspace{2mm}
\caption{$T$-dependent $\protect\sigma (\protect\omega )$ (a) above and (b)
below $T_{C}$. At $T_{C}<T$, the optical gap energy due to the CO phase is
determined by drawing a linearly extrapolated line (dotted) at the
inflection point of $\protect\sigma (\protect\omega )$. At $T\leq $ $T_{C}$,
at least two absorption bands appear in the mid-infrared region. The peak
positions of the bands are indicated as asterisks.}
\label{Fig:3}
\end{figure}

\begin{figure}[tbp]
\epsfig{file=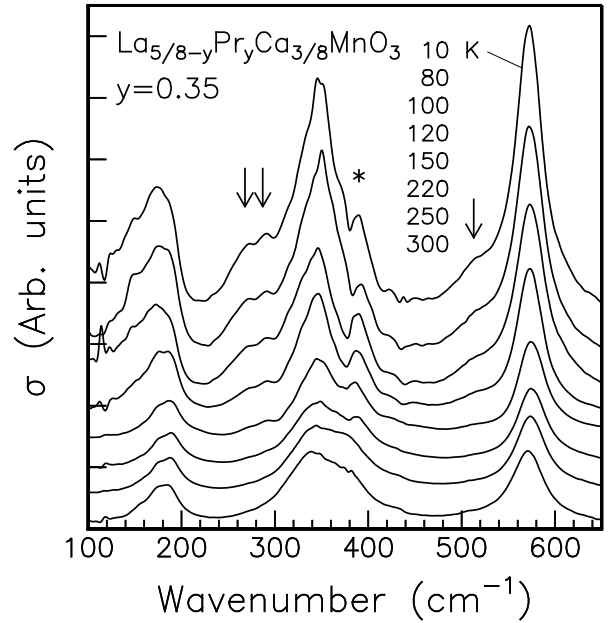, width=3.1in,clip=}
\vspace{2mm}
\caption{Optic phonon modes of La$_{5/8-y}$Pr$_{y}$Ca$_{3/8}$MnO$_{3}$ ({\it %
y}=0.35). Arrows indicate the additional phonon modes appearing below $T_{%
\text{CO}}$.}
\label{Fig:4}
\end{figure}

 mode is sensitive to
the Mn-O bond length. As shown in Fig. 4, the bending mode starts to be
split near {\it T}$_{\text{CO}}$. Similar bending mode splitting{\em \ }was
often observed when large anisotropic lattice distortions developed with the
stabilization of charge ordering.\cite{katsu} As {\it T} decreases below 
{\it T}$_{\text{C}}$, the additional bands around 270, 290 and 516 cm$^{-1}$
seem to continuously grow in intensity with{\em \ }the bending mode
splitting. The persistence of the bending mode splitting below {\it T}$_{%
\text{C}}$ indicates that the CO phase still remains in the low {\it T}.

\subsection{{\it H}-dependence of optical spectra}

In order to further understandings, we performed melting experiments for the
low-temperature CO phase by applying a magnetic field. As shown in Fig. 5,
the {\it H}-dependent $R(\omega )$ were measured at a fixed {\it T} of 4.2 K
using the facilities at{\em \ }National High Magentic Field Laboratory at
Tallahassee. Details of the {\it H}-dependent reflectivity measurements were
described elsewhere.\cite{jungH} With increasing {\it H}, the reflectivity
increases and the phonon modes become screened. The inset of Fig. 5 shows
the hysteresis of the $\rho $ values measured at 4.2 K. With increasing {\it %
H}, the $\rho $ value decreases abruptly at low {\it H} below 4 T and nearly
saturates above 4 T. With decreasing {\it H} from 12 T to 0 T, $\rho $
remains at a small value, not recovering to the original value at 0 T. This
hysteresis can be explained by the fact that the CO phase melts into a FM
phase in the field increasing run, but the sample remains in the FM phase in
the field decreasing run. \ Similar metastable FM phases were reported in Pr$%
_{0.69}$Ca$_{0.31}$MnO$_{3}$ and (La$_{0.7}$,Nd$_{0.3}$)$_{1.4}$Sr$_{1.6}$Mn$%
_{2}$O$_{7}$.\cite{HJLee2}

\begin{figure}[tbp]
\epsfig{file=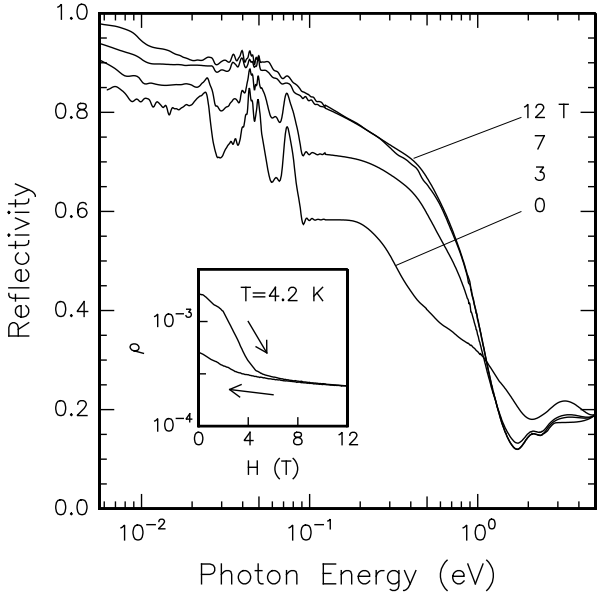, width=3.1in,clip=}
\vspace{2mm}
\caption{{\it H}-dependent R($\protect\omega $) of La$_{5/8-y}$Pr$_{y}$Ca$%
_{3/8}$MnO$_{3}$ ($y$=0.35) single crystal at 4.2 K. Inset : {\it H}%
-dependent $\protect\rho $ measured at 4.2 K.}
\label{Fig:5}
\end{figure}

Figure 6 shows the {\it H}-dependent $\sigma (\omega )$, which were obtained
from the KK analysis of $R(\omega )$ in Fig. 5. With increasing {\it H}, the
spectral weight above 1.0 eV becomes strongly suppressed and transferred to
a low frequency region. Although the sample shows a metallic resistivity at
3 T, the corresponding optical spectrum still has an asymmetric mid-IR band.
At {\it H}=12 T, a Drude-like peak can be clearly observed and the
absorption peak around 1.4 eV disappears. Strong asymmetric absorption band
below 1.4 eV were observed in the $\sigma (\omega )$ of some homogeneous FM
metallic manganites and attributed to the large polaron absorption.\cite
{KHKim98} Note that the two absorption bands shown in Fig. 3(b), are not
observed in the data of Fig. 6 under high{\it \ H}.{\em \ }Therefore, it is
evident that the optical spectra in Fig. 3(b) cannot be interpreted in terms
of responses of a homogeneous FM metal.

\begin{figure}[tbp]
\epsfig{file=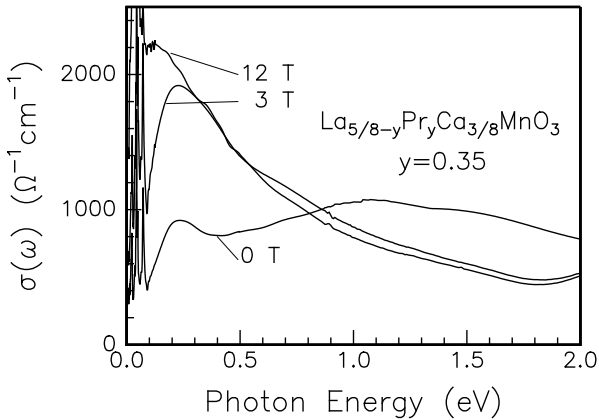, width=3.1in,clip=}
\vspace{2mm}
\caption{{\it H}-dependent $\protect\sigma (\protect\omega )$.}
\label{Fig:6}
\end{figure}

\section{DISCUSSION}

\subsection{Failure of two phase description{\em \ }}

One important question which we should address at this point is how we can
interpret the low-frequency absorption features below {\it T}$_{\text{C}}$.
In Fig. 7(a), we plot together the {\it T}-dependent $\sigma (\omega )$ at
zero field and the {\it H-}dependent $\sigma (\omega )$ at fixed 4.2 K. Note
that the $\sigma (\omega )$ with {\it H=}0 T and {\it T}=150 K represents
the CO response and the $\sigma (\omega )$ with {\it H=}12 T and {\it T}=4.2
K represents the FM response. With decreasing {\it T} from 150 K to 4.2 K,
the spectral weight below 1 eV increases. And, with increasing {\it H} at
4.2 K, a large spectral weight transfer occurs from above $\sim $1 eV to
below $\sim $1 eV. Then, the simplest way to describe $\sigma (\omega )$
below {\it T}$_{\text{C}}$ is to use two phase{\em \ }description [namely,
by assuming that the each state is composed of the insulating CO and the metallic FM phases].

Optical properties of an inhomogeneous medium can be modeled by several
effective medium theories, which predict the effective optical conductivity $%
\sigma _{eff}(\omega )$ in terms of $\sigma (\omega )$ and volume fraction
of its constituent components. To check the validity of the two phase
description, we applied the two most commonly used effective medium
theories, i.e., the Maxwell-Garnet theory (MGT) and the Bruggeman-type
effective medium approximation (EMA). Details of these theories and their
application to optical properties of composites can be found elsewhere.\cite
{Lee} Figures 7(b) and 7(c) show $\sigma _{eff}(\omega )$ calculated using
the MGT and the EMA for various values of $f$, which represents the volume
fraction of the metallic domains, respectively. We used $\sigma (\omega )$
at 150 K and at 12 T (and 4.2 K) as responses of the homogeneous insulating
and metallic phases, respectively. It looks like that the MGT is more
appropriate for describing the general features of the experimental data,
shown in Fig. 7 (a). However, none of the theories could reproduce the two
peak structure below 0.5 eV, which was marked with the asterisks. This shows
that a simple picture based on the two phase description{\em \ }cannot
explain the two absorption features observed below {\it T}$_{\text{C}}$.

\subsection{Optical Evidences of multiphase coexistence}

To obtain further insight, we subtracted the optical response of the CO
phases from the measured $\sigma $($\omega $) at each temperature, $\sigma $(%
$\omega ,${\it T}). It was assumed that $\sigma $($\omega ,$150 K) could
represent the $\sigma $($\omega $) of CO domains. Figure 8 shows the results
for $\Delta \sigma $($\omega ,${\it T})$\equiv \sigma $($\omega ,${\it T})-$%
\sigma $($\omega ,$150 K) at various temperatures. The $\Delta \sigma $($%
\omega ,$10 K) curve is composed of an asymmetric absorption band peaked
around 0.2 eV and a broad band with peaks around 0.4 and 0.8 eV. While the
absorption band around 0.2 eV appears below {\it T}$_{\text{C}}$, the broad
band with peaks around 0.4 and 0.8 eV already exists above {\it T}$_{\text{CO%
}}$.

\begin{figure}[tbp]
\epsfig{file=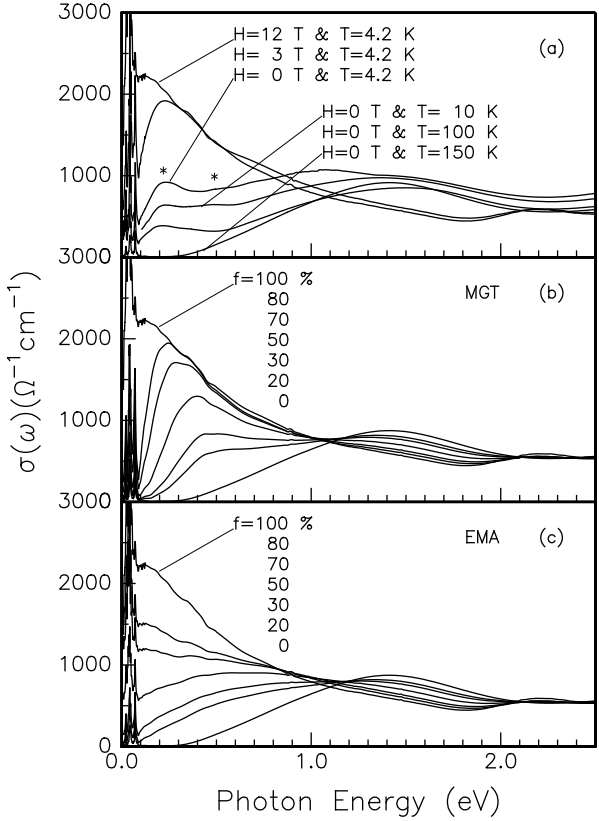, width=3.1in,clip=}
\vspace{2mm}
\caption{(a) $T$- and $H$-dependent $\protect\sigma (\protect\omega )$. (b)
The MGT and (c) the EMA predictions with various metal volume fractions $f$.}
\label{Fig:7}
\end{figure}

To quantitatively estimate these {\it T}-dependent spectral weight
absorption bands, we fitted the $\Delta \sigma $($\omega ,${\it T}) below 
{\it T}$_{\text{C}}$ as a sum of an asymmetric band (Band I) around 0.2 eV
and two Lorentzians (Band II) around 0.4 and 0.8 eV. The inset of Fig. 8
shows the $\Delta \sigma $($\omega ,$10 K) curve and its fitting results. It
is found that Band I is very similar to $\sigma (\omega )$ at {\it H}=3 T in
Fig. 6. Band II is very similar to $\Delta \sigma $($\omega ,$280 K) in Fig.
8. These observations strongly suggest that Band I at low {\it T} should be
associated with the FM metallic phase and that Band II should be attributed
to another phase. The physical properties of this additional phase can be
very similar to those of the high {\it T} charge-disordered (CDO) insulating
phase.

Based on these analyses, we estimated the strength of Band I, S$_{I}$, and
Band II, S$_{II}$, below {\it T}$_{\text{C}}$. It is interesting to compare
the {\it T-}dependence of S$_{I}$ and S$_{II}$ with the normalized
magnetization value, M({\it T})/M(0). Figure 9(a) shows {\it T}-dependences
of S$_{I}$ (solid squares), S$_{II}$ (solid triangles), and M({\it T})/M(0).
The observation that S$_{I}$ is proportional to M({\it T})/M(0) indicates
that Band I is a result of the FM spin ordering. And the smooth behavior of
M({\it T})/M(0) is not consistent with the FM transition in a homogeneous
system. Therefore, S$_{I}$ can be attributed to the spectral weight of FM
metallic domains in an inhomogeneous system. On the other hand, S$_{II}$
increases with M({\it T})/M(0) near {\it T}$_{\text{C}}$ and it

\begin{figure}[tbp]
\epsfig{file=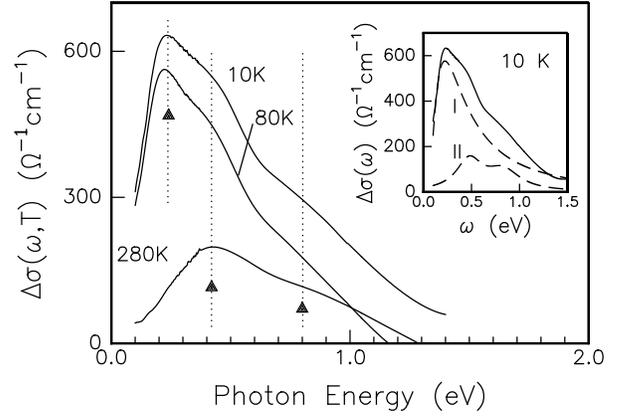, width=3.1in,clip=}
\vspace{2mm}
\caption{$\Delta \protect\sigma $($\protect\omega ,T$)$\equiv \protect\sigma 
$($\protect\omega ,T$)-$\protect\sigma $($\protect\omega ,$150 K) for
various temperatures. The $\Delta \protect\sigma $($\protect\omega ,T$)
curves at $T$%
\mbox{$<$}%
\mbox{$<$}{\it T}$_{\text{C}}$ are composed of an asymmetric absorption band
and small additional bands (filled triangles). The spectral shape of the
additional bands are very similar to the shape of $\Delta \protect\sigma $($%
\protect\omega ,T$) at $T$%
\mbox{$>$}%
\mbox{$>$}{\it T}$_{\text{CO}}$. The inset shows a $\Delta \protect\sigma $($%
\protect\omega ,$10 K) curve and fitting results using an asymmetric line
shape (for Band I) and two Lorentzians (for Band II).}
\label{Fig:8}
\end{figure}

continuously
increases even when M({\it T})/M(0) is saturated. This again confirms the
above conclusion that Band II is due to the absorption of a CDO insulating
phase appearing additionally with the development of the FM phase below {\it %
T}$_{\text{C}}$. {\it All of these experimental findings suggest that there
exist at least three phases, namely the FM metallic, the CO insulating, and
the CDO insulating phases in LPCMO.}

\subsection{A possible origin of the charge disordered phase}

The existence of the CDO phase in LPCMO might be closely related to the
strain developments suggested by Littlewood.\cite{Littlewood} In perovskite
manganites with strong Jahn-Teller (JT) electron-phonon interaction,
anisotropic lattice strain can be developed by the JT distortion. In the CO
state, it is well known that the anisotropic strain is quite large due to a
cooperative JT distortion and a {\it d}$_{z^{2}}$ orbtial ordering. On the
other hand, in a nearly homogeneous FM metallic state, the JT distortion
becomes small. As {\it T} decreases below {\it T}$_{\text{C}}$, the FM phase
starts to grow. If a single FM crystallite nucleates into the CO phase, it
will be under a large stress from the surrounding CO crystals that
discourages further growth. Then, domains in different parts of the LPCMO
crystal will form with the strain field in random orientations, so the
inhomogeneous strain cannot be easily released, leading to phase separation
with a sub-$\mu $m length scale. This can explain the dominance of the CO
phases far below {\it T}$_{\text{C}}$ and the hysteretic behavior of $\rho $%
, shown in the inset of Fig. 5. Moreover, due to a large strain mismatch
between the FM and the CO phases, the interfacial region will have
inhomogeneous strain and form the CDO phase.

\begin{figure}[tbp]
\epsfig{file=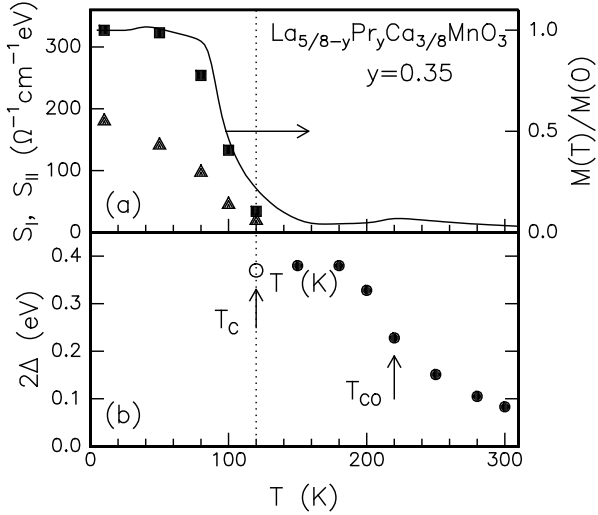, width=3.1in,clip=}
\vspace{2mm}
\caption{(a) $T$-dependence of spectral weight S$_{I}$ (solid squres) and S$%
_{II}$ (solid triangles) at $T\leq $ $T_{C}$. See inset of Fig. 4 and texts
for definition. A solid line represents a normalized magnetization curve.
(b) $T$-dependence of the optical gap energy 2$\Delta $ (solid circles). An
open circle represents a 2$\Delta $ at $T_{C}$. }
\label{Fig:9}
\end{figure}

The temperature dependence of the phonon spectra, shown in Fig. 4, also
supports the suggested scenario based on the strain developments. It is well
known that the phonon modes are quite sensitive to local lattice
distortions. Figure 4 shows that, with decreasing {\it T}, the phonon mode
splitting around 346 cm$^{-1}$ starts to occur near {\it T}$_{\text{CO}}$
and remains even at {\it T}%
\mbox{$<$}%
{\it T}$_{\text{C}}$, in agreement with the dominance of the CO phase at the
low temperature. Moreover, a few additional phonon structures, which are
marked with arrows, clearly appear below {\it T}$_{\text{CO}}$, consistent
with the existence of a new phase (i.e., the CDO phase).

The asymmetric line shape of Band I is indicative of the fact that the
lattice distortion of the FM domains is not large.\cite{comment} On the
other hand, the $\sigma (\omega )$ of the CO domains with large strains
showed a band centered $\sim $1.4 eV with a large 2$\Delta \approx $0.4 eV.
Most of the spectral weights of Band II appear in the energy region of 0.3
-1.0 eV. This indicates that the lattice strain of the CDO region can be
larger than that of the FM metallic domains, but smaller than that of the CO
domains. According to our results in Fig. 3(b) and Fig. 8(a), the lower
center frequency of Band II and S$_{II}$ increased as {\it T} decreased
below {\it T}$_{\text{C}}$. This seems to indicate that the volume fraction,
as well as the strain of the CDO domains, possibly located at the interface
of FM and CO domains increases with deceasing {\it T}. Therefore, our $%
\sigma (\omega )$ data suggest that the lattice strains and their interplay
with {\it T} among the three phases play a crucial role in determining the
electronic and the magnetic properties of the LPCMO.

\subsection{Charge ordering gap of the CE-type CO phase}

It is worthwhile to acquire quantitative information on the optical gap
energy 2$\Delta $ at {\it T}$_{\text{C}}\leq ${\it T}. We drew a linear
tangential line at the inflection point of $\sigma (\omega )$ and assigned
its crossing energy with the abscissa as the gap value, shown as a dotted
line in Fig. 3(a). Figure 9(b) shows the 2$\Delta $ vs. {\it T} plot. The 2$%
\Delta $ just above {\it T}$_{\text{C}}$ is found to be as large as 0.38 eV.
The value remains nearly the same for {\it T}$_{\text{C}}\leq ${\it T}$\leq $%
180 K and decreases slightly near {\it T}$_{\text{CO}}$. It should be noted
that the 2$\Delta $ $\approx $ 0.38 eV at {\it T}$\approx $150 K is
comparable to the observed value of La$_{1/2}$Ca$_{1/2}$MnO$_{3}$ (i.e., 2$%
\Delta (0)$ $\approx $ 0.45 eV at the ground state).\cite{KHKim2} Since the
charge-ordering in LPCMO is known to be of the La$_{1/2}$Ca$_{1/2}$MnO$_{3}$%
-type, i.e., the so-called CE-type,\cite{Uehara} the large value of 2$\Delta 
$ of LPCMO can be ascribed to the characteristics of the CE-type CO phase.

It is also interesting to note that 2$\Delta $ $\approx $0.22 eV at {\it T}$%
_{\text{CO}}$ and 2$\Delta \approx $0.1 eV at 300 K. In other words, 2$%
\Delta $ does not become zero at {\it T} far above {\it T}$_{\text{CO}}$.
This is in contrast with the behavior of many CO materials that show a
nearly zero value of 2$\Delta $ at {\it T}$_{\text{CO}}$. These anomalous 2$%
\Delta $ behaviors indicate that there exist enhanced spatial and/or
temporal fluctuations in the CO correlation far above {\it T}$_{\text{CO}}$
in LPCMO.\cite{KHKim2} The enhanced CO fluctuations can be a generic feature
of LPCMO that has mixed-phases near the phase boundary where a phase
separation occurs as {\it T}$\rightarrow $0.\cite{Moreo}

\section{SUMMARY}

In conclusion, {\it T}- and {\it H}-dependent optical conductivity spectra
of a La$_{\text{5/8}-y}$Pr$_{y}$Ca$_{\text{3/8}}$MnO$_{\text{3}}$ single
crystal revealed that at {\it T}%
\mbox{$<$}%
{\it T}$_{\text{C}}$, at least two absorption bands newly emerge below 0.4
eV. The new absorption bands can be attributed to a FM metallic and a
charge-disordered phase that coexist with the charge-ordered phase. We also
found that the LPCMO has a rather large charge gap and shows a fluctuation
in charge-ordered correlation above {\it T}$_{\text{C}}$. In addition, the 
{\it T}-dependent changes of the phonon modes as well as polaron bands
suggest that the coexisting multiphase might have different lattice strains.
Our $\sigma (\omega )$ study supports that the structural as well as the
electronic phase separation occurs in the LPCMO below {\it T}$_{\text{C}}$.

\section{ACKNOWLEDGMENTS}

We are grateful to H. K. Lee, Dr. J. H. Jung, Y. S. Lee, D. S. Suh and Dr.
Y. Chung for useful discussions and helpful experiments. This work was
supported by the Ministry of Science and Technology through the Creative
Research Initiative Program and by the Ministry of Education through the BK
21 Program. BGK, TYK and SWC are partially supported by NSF-DMR-9802513 and
NSF-DMR-00-80008. Part of this work was performed at the National High
Magnetic Field Laboratory, which is supported by the NSF Cooperative
Agreement No. DMR-0084173 and by the State of Florida.

\end{document}